\documentclass[prc,preprint]{revtex4}
\usepackage{graphicx}
\usepackage{amssymb}

\begin{document}

\title{\bf Influence of strong magnetic field on the structure properties of strange quark stars}

\author{{ Fatemeh Kayanikhoo$^{1}$
}, { Kazem Naficy $^{1}$} and {Gholam Hossein Bordbar  $^{2,3}$}\footnote{Corresponding author. E-mail:
ghbordbar@shirazu.ac.ir}}
\affiliation{$^1$ Department of Physics, University of Birjand, Birjand, Iran
\\ $^2$ Department of Physics,
Shiraz University, Shiraz 71454, Iran\footnote{Permanent address}\\
and\\  $^3$ Department of Physics and Astronomy, University of Waterloo,
200 University Avenue West, Waterloo, Ontario,
N2L 3G1,
Canada
}
\begin{abstract}

We investigate the thermodynamic properties of strange quark matter under the strong magnetic field in the framework of the MIT bag model in two cases of bag constants.
We consider two cases of the magnetic field, the uniform magnetic field and the density-dependent magnetic field to calculate the equation of state of strange quark matter.
For the case of density-dependent magnetic field, we use a Gaussian equation with two free parameters $\beta$ and $\theta$ and use two different sets of the parameters for the magnetic field changes  (a slow and a fast decrease of the magnetic field from the center to the surface).
Our results show that the energy conditions based on the limitation of the energy-momentum tensor, are satisfied in the corresponding conditions.
We also show that the equation of state of strange quark matter becomes stiffer by increasing the magnetic field.
In this paper, we also calculate the structure parameters of a pure strange quark star using the equation of state.
{We investigate the compactification factor ($2M/R$) and the surface redshift of star in different conditions.}
The results show that the strange quark star is denser than the neutron star and it is more compact in the presence of the stronger magnetic field.
As another result, the compactification factor increases when we use a slow increase of the magnetic field from the surface to the center.
Eventually, we compare our results with the observational results for some strange star candidates, and we find that the structure of the strange star candidates is comparable to that of the star in our model.

\end{abstract}

\maketitle

\section{Introduction}

Like white dwarfs, neutron stars, and black holes, strange quark stars (SQS) are compact objects, created at the end of the life of the massive stars.
The SQS is denser than the neutron star. In other words, they have the maximum gravitational masses around $1.7-2 M_{sun}$ as the neutron stars but the smaller radii (around $4-8\ km$) \cite{rk1}.
The SQS is considered in two cases: 1) The pure SQS that all over is made of strange quark matter (SQM) 2) The hybrid star that is the neutron star with the core made of SQM.

When the nuclear matter is compressed to the high density (more than $10^{15} g/cm^{3}$), the nucleons overlap and it is expected that the nuclear matter converts to the quark matter
(consisting of up and down quarks) by phase transition \cite{rk2, rk3}. The weak interaction between up and down quarks creates strange quarks \cite{rk4}.
For the first time, Gell-Mann and Zweig suggested that hadrons are made up of the smaller particles later called quarks\cite{rk5, rk6}. Ivanenco explained the possibility of the existence of a quark core in some compact objects \cite{rk7, rk8}.
In 1970, Itoh computed the maximum gravitational mass of the quark stars \cite{rk9}. In addition, in 1984, Witten suggested that the basic matter of the stars could be the strange matter \cite{rk10}.
There have been candidates for quark stars, RX J185635-3754 and 3C58 observed by Chandra X-ray Observatory in 2002 \cite{rk1}, and SWIFT J1749.42807 \cite{rk11}, etc.

The most important properties of compact objects is a strong magnetic field. The surface magnetic field that observed for white dwarfs is about $10^{6}-10^{8}\ G$ \cite{rk12}, and for the neutron stars and SQS's is about $10^{12}-10^{14}\ G$ \cite{rk13}.
It is clear that the interior magnetic field of compact objects is a few orders of magnitudes stronger. So the interior magnetic field around $10^{12}$-$10^{14}\ G$  for the white dwarfs and $10^{18}-10^{20}\ G$ for the neutron stars and SQS's are expected \cite{rk13, rk14, rk15}.
Although the source of these strong magnetic fields is not clear, it can be considered that existence of a large number of spin particles creates a strong magnetic flux and as a result a strong magnetic field.
In presence of the strong magnetic field, the Landau quantization effect is considerable. In other words, the strong magnetic field causes the cyclotron orbits of the charged particles to be quantized \cite{rk16, rk17}.
Several papers have already been published on the magnetic field effect on compact objects. Mukhopadhyay investigated the Landau quantization effect on white dwarfs \cite{rk12}. Rezaei and Bordbar published papers in subjects of the effect of Landau quantization
on neutron matter \cite{rk18,rk19}. In addition, Lai and Shapiro published a paper on the same subject \cite{rk13}. Lopes and Menezes considered the Landau quantization effect in neutron stars \cite{rk20}.

Bordbar et. al. published some papers on the SQS's in the various conditions in recent years. They investigated the thermodynamic properties and structure of cold polarized SQS \cite{rk21,rk22} and the magnetized SQS at the finite temperature \cite{rk23,rk24}. In addition, Non-polarized SQS using MIT bag model and NJL model have been investigated \cite{rk25, rk26, rk27}. Furthermore, papers have been published to investigate properties of neutron stars with the quark core \cite{rk28,rk29}.

In the current paper, we investigate the effect of the strong magnetic field on SQS with Landau quantization effect using the MIT bag model in two cases of bag constants. {It should be mentioned that in our calculations, we consider an isotropic magnetic field for the interior of SQS}.  In the next section, we study the formalism of calculation of the thermodynamic properties of SQM. The structure parameters of SQS are calculated in the third section. The conclusion is given in the final section.

\section{Formalism of Calculation }
\label{I}
In this section, we investigate the thermodynamic properties of a pure SQS made by up, down and strange quarks. Since the density of electrons is low compared to that of quarks, it is ignored \cite{rk30,rk31,rk32}.
The SQM is a relativistic Fermi system, so we should use Fermi-Dirac statistics to calculate its thermodynamic properties \cite{rk33}.
We use the MIT bag model to calculate the energy density of the system. The MIT bag model is an approach based on QCD model, in this model quarks are considered as the free particles in a bag under the potential of a bag constant \cite{rk34,rk35}.

\subsection{Energy density and equation of state of strange quark matter}
To study the thermodynamic properties and structure of SQS, the first step is to calculate the energy density and the equation of state (EOS).
As we mentioned, we use the Fermi-Dirac statistics and consider the Landau quantization effect to calculate the thermodynamic properties of SQM in the presence of a strong magnetic field.
By considering the Landau quantization effect the single particle energy of $J$th Landau levels is {(see Refs. \cite{rk12,rk13,rk16,rk17} for more details)},
\begin{equation}\label{01}
\epsilon^{i}=[p_{i}^{2}c^{2}+m_{i}^{2}c^{4}(1+2JB_{D})]^{1/2},
\end{equation}
where superscript $i$ represents the particles ($i=$ up, down, strange quarks), {$p$, $c$ and $m$ are momentum, speed of  light and mass of particles, respectively}, and $B_{D}=B/B_{c}$ ($B$ is the magnetic field and $B_{c}$ is equal to $m_{i}^{2}c^{3}/q_{i}\hbar$ {where $q$ is charge of particles and $\hbar$ is Plank constant}).
The number density of quarks is obtained as follows,
\begin{equation}\label{02}
\rho=\sum_{J=0}^{J_{max}} \frac{2qB}{h^{2}c}g_{J}p_{F}(J) ,
\end{equation}
where $p_{F}(J)$ is the Fermi momentum corresponding to the $J$th. Landau level, $g_{J}$ is the degeneracy of the $J$th. Landau level and $J_{max}$ is the upper limit of Landau level,
\begin{equation}\label{03}
J_{max}=\frac{\epsilon_{Fmax}^{2}-1}{2B_{D}}
\end{equation}
in the above equation, $\epsilon_{Fmax}$ is the Fermi energy density corresponding to the Landau level.

In the frame work of MIT bag model, the total energy density of the system is,
\begin{equation}\label{04}
\varepsilon_{tot}=\sum_{j=+,-}\varepsilon_{i}^{(j)}+\varepsilon_{M}+B_{bag},
\end{equation}
in the above equation, we denote the bag constant as $B_{bag}$, we consider two cases for $B_{bag}$, a fixed bag constant and a density-dependent bag constant. For the fixed bag constant we use $B_{bag}^{fixed}=90 MeV/fm^{3}$,
and for the density-dependent bag constant, we use the Gaussian equation given by experimental data of CERN as follows,
\begin{equation}\label{10}
B_{bag}^{dep.}=B_{bag}(\rho)=B_{\infty}+(B_{0}-B_{\infty})exp{(-\beta (\frac{\rho}{\rho _{0}})^{2})},
\end{equation}
where $\beta$ is the numerical parameter equal to the density of normal nuclear matter ($\rho_{0}=0.17\ fm^{-3}$), and $B_{0}=B(\rho= 0)$ is equal to $400\ MeV/fm^{3}$ {\cite{rk36a,rk36b,rk36,rk37}}.
Also, the value of the parameter $B_{\infty}$   depends only on the value of the parameter $B_{0}$ and is obtained by the LOCV method in our calculations \cite{rk38,rk39,rk40,rk41}.
In Eq. \ref{04} the energy density due to the interaction of the magnetic field and the dipole moment of the quarks is denoted
by $\varepsilon_{M}$ ($\varepsilon_{M}=0.299\rho \xi \mu_{N} B$, $\mu_{N}$ is $5.05\times 10^{-27}$$J/T$ and $\xi$ is the polarization parameter of system $\xi=\frac{\rho^{(+)}-\rho^{(-)}}{\rho}$)
and the kinetic energy density of quarks is denoted by $\varepsilon_{i}^{(j)}$ ($j=+,-$, for spin up and  spin down particles, respectively).
The kinetic energy density of each quark $\varepsilon_{i}^{(j)}$ is as follows {\cite{rk33}},
\begin{equation}\label{05}
\varepsilon_{i}^{(j)}=\frac{2B_{D}}{(2\pi)^{2}\lambda^{3}}m_{i}c^{2}\sum_{J=0}^{J_{max}} g_{J}(1+2JB_{D})\eta(\frac{X_{F}^{(j)}}{(1+2JB_{D})^{1/2}}),
\end{equation}
where
\begin{equation}\label{06}
\eta(x)=\frac{1}{2}[x\sqrt{1+x^{2}}+ln(x\sqrt{1+x^{2}})], x=\frac{X^{(j)}_{F}}{(1+2JB_{D})^{1/2}}
\end{equation}
\begin{equation}\label{07}
X^{(j)}_{F}=(\epsilon_{F}^{(j)2}-1-2JB_{D})^{1/2}.
\end{equation}

The equation of state (EOS) of SQM is given by the following relation,
\begin{equation}\label{09}
P(\rho)=\rho (\frac{\partial \varepsilon_{tot}}{\partial \rho}) - \varepsilon_{tot}.
\end{equation}

\subsection{Magnetic energy density and magnetic pressure}
\label{B}
As there is a strong magnetic field, the magnetic energy density and magnetic pressure cannot be ignored.
The magnetic energy density $\varepsilon_{B}=\frac{B^2}{8\pi}$ and the magnetic pressure $P_{B}=\frac{B^2}{8\pi}$ ($B$ is the magnetic field) are directly added to the total energy density ($\varepsilon_{tot}$) and the EOS ($P(\rho)$) \cite{rk42,rk43,rk44,rk45}.

Here we study the uniform magnetic field and the density-dependent magnetic field. In the first case, $B$ is constant from the center to the surface of SQS. We compare the different magnetic fields effects in the section of calculation results.
In the second case, the magnetic field is considered as a function of the density of the system. We use a Gaussian function as follows,
\begin{equation}\label{11}
B(\rho)=B_{surf}+B_{0}[1-exp{\beta (\frac{\rho}{\rho_{0}})^{\theta}}],
\end{equation}
where $B_{surf}$ is the magnetic field of the surface of the SQS that we consider $5\times 10^{13}\ G$ and $B_{0}$ is the interior magnetic field that is expected for the star. Also, $\beta$ and $\theta$ are the
parameters that define the magnetic field changes based on the density of SQS \cite{rk42,rk46,rk47}.
In other words, these parameters select as the magnetic field decrease fast or slow from the center to the surface of the star. There are different sets of $\beta$ and $\theta$ investigated in the literature \cite{rk48, rk49}.

\subsection{Thermodynamic calculation results}
\label{results}
We have shown the pressure of the system versus the total energy density in Figs. \ref{01} and \ref{02} by considering $B_{bag}^{fixed}$ and $B_{bag}^{dep.}$, respectively.
The figures are plotted in the case of the uniform magnetic fields and the effect of the different magnetic fields are compared.
{The figures show that pressure increases by increasing the energy density in all cases. Furthermore, The EOS becomes stiffer by increasing the magnetic field.
Also, It can be seen that the energy density in the case of strong magnetic fields is much greater than the case with no magnetic field ($B = 0$).
The Landau quantization effect breaks the spherical symmetry and thereby increases the energy density up to several times of the normal nuclear energy density.}
\begin{figure}[h]
\centering
\includegraphics[width=10cm, height=7cm]{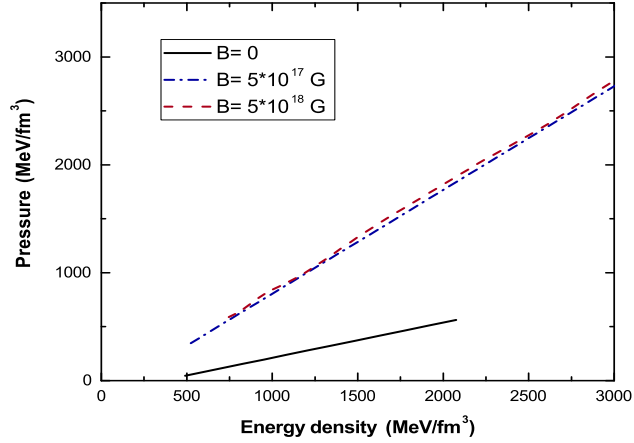}
\caption{The pressure versus the energy density of system in presence of different magnetic fields using $B_{bag}^{fixed}= 90\ MeV/fm^{3}$. } \label{01}
\end{figure}
\begin{figure}[h]
\begin{center}$
\begin{array}{cc}
\includegraphics[width=9cm, height=6.3cm]{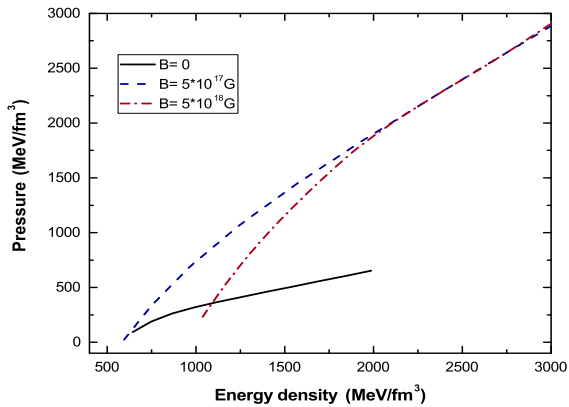}
\end{array}$
\end{center}
\caption{The pressure versus the energy density of system in presence of different magnetic fields using $B_{bag}^{dep.}$.} \label{02}
\end{figure}
In Figs. \ref{04} and \ref{05} the case of the density-dependent magnetic field have been plotted by considering $B_{bag}^{fixed}$ and $B_{bag}^{dep.}$, respectively. We considered two different sets of $\beta$ and $\theta$:
The fast changes of the magnetic field (Set A: $\beta= 0.05 $ and $\theta= 2$) and the slow changes of the magnetic field (Set B: $\beta= 0.1 $ and $\theta= 1$).
It can be seen from Fig. \ref{04} that considering the slow changes of the magnetic field leads to the softer equation of state in comparison to the fast changes of that. The same behavior can be seen in Fig. \ref{05} for the case of density-dependent bag constant.
\begin{figure}[h]
\centering
\includegraphics[width=10cm, height=7cm]{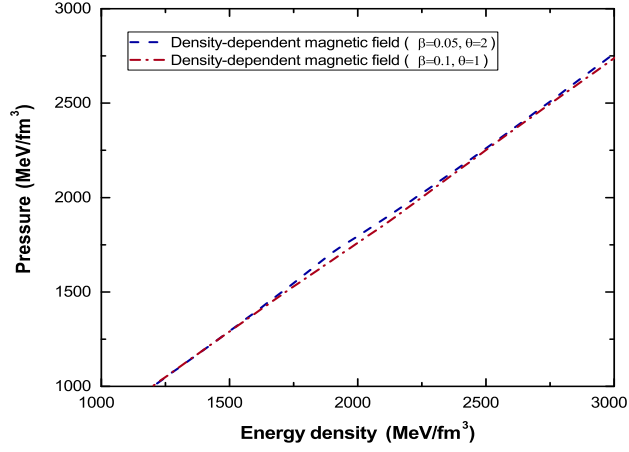}
\caption{The pressure versus the energy density of system in presence of the case of density-dependent magnetic field (Set A and Set B) using $B_{bag}^{fixed}= 90\ MeV/fm^{3}$. } \label{04}
\end{figure}
\begin{figure}[h]
\centering
\includegraphics[width=10cm, height=7cm]{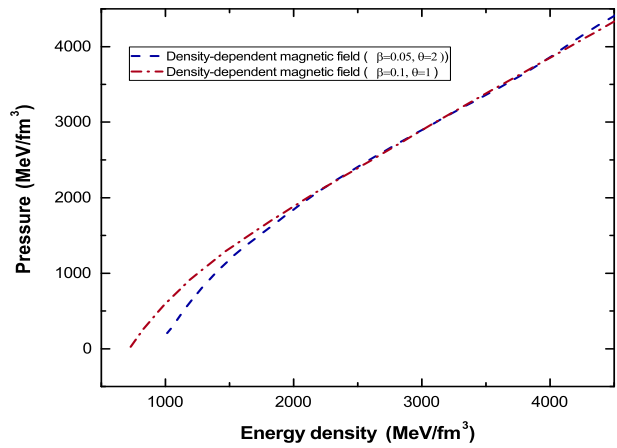}
\caption{The pressure versus the energy density of system in presence of the case of  density-dependent magnetic field (Set A and Set B) using $B_{bag}^{dep.}$. } \label{05}
\end{figure}
We know that any theory of physics must comply with the energy conditions that result from the limitation in energy-momentum tensor, $T^{\mu \nu}$ \cite{rk48,rk49}.
We can ensure the accuracy of the results obtained for the energy and the pressure of the system by investigation the energy conditions:
{
\\
a) Null energy condition (NEC) $\longrightarrow$ $P_{c}$ + $\rho_{c} c^{2}$$ \geq$0,\\
b) Weak energy condition (WEC)  $\longrightarrow$ $P_{c}$ + $\rho_{c} c^{2}$$ \geq$0 and  $\rho_{c}$$ \geq$ 0,\\
c) Strong energy condition (SEC) $\longrightarrow$ $P_{c}$ + $\rho_{c} c^{2}$$ \geq$ 0 and $3P_{c}$ + $\rho_{c} c^{2}$$ \geq$ 0,\\
d) Dominate energy condition (DEC) $\longrightarrow$ $ \rho_{c} c^{2}$$ \geq $$\mid$ $P_{c}$ $\mid$,\\
\\
where $\rho_{c}$ and $P_{c}$ are the mass density and pressure at the center of SQS ($r=0$).}
We have found that the energy conditions are satisfied regarding the equation of state for all considered cases of this paper.

\section{Structure properties of strange quark star}
\label{struc}
The structure of stars is usually determined by their mass, radius and some other parameters.
In this section, we study the structure of the SQS by gravitational mass
($M/M_{sun}$), radius ($R$), compactification factor ($2M/R$)and surface redshift ($Z_{s}$).
For the compact objects, we should use the general relativistic equation of hydrostatic equilibrium known as Tolman-Oppenheimer-Volkov (TOV) equations \cite{rk50, rk51,rk52},
\begin{equation}\label{10}
\frac{dP}{dr}=-\frac{G\left[\varepsilon(r)+\frac{P(r)}{c^{2}}\right]\left[m(r)+\frac{4\pi
r^{3}P(r)}
{c^{2}}\right]}{r^{2}\left[1-\frac{2Gm(r)}{rc^{2}}\right]},
\end{equation}
\begin{equation}\label{11}
\frac{dm}{dr}=4\pi r^{2}\varepsilon(r).
\end{equation}
Using the thermodynamic results of the previous section and numerical solution of TOV equations the gravitational mass and the radius of the star have been computed.
We consider the usual boundary conditions: $P(r=0)=P_{c}$, $P(r=R)=0$, $m(r=0)=0$ and $m(r=R)=M_{max}$.
\subsection*{Results for calculations of the structure}
We have shown the gravitational mass versus the central energy density of the SQS in Figs. \ref{06} and \ref{07} for the cases of $B_{bag}^{fixed}$ and $B_{bag}^{dep.}$, respectively.
In these figures, we compared the behavior of the gravitational mass of SQS in the presence of different magnetic fields.
These figures show that for all curves the gravitational mass increases by increasing the central energy density until it reaches a maximum value (The maximum gravitational mass).
Also, it can be seen that the maximum gravitational mass of SQS has a larger value in the absence of the magnetic field.
\begin{figure}[ht]
\centering
\includegraphics[width=10cm, height=7cm]{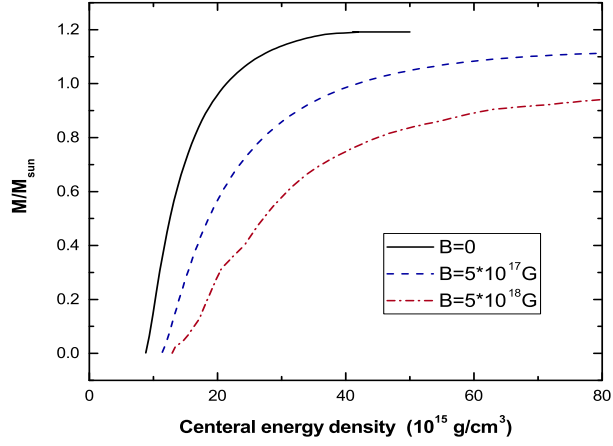}
\caption{The gravitational mass versus the central energy density of SQS in presence of different magnetic fields using $B_{bag}^{fixed}= 90\ MeV/fm^{3}$. } \label{06}
\end{figure}
\begin{figure}[ht]
\centering
\includegraphics[width=10cm, height=7cm]{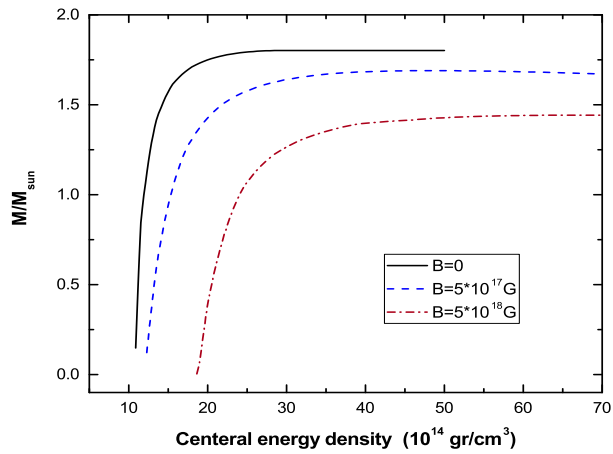}
\caption{The gravitational mass versus the central energy density of SQS in presence of different magnetic fields using $B_{bag}^{dep.}$. } \label{07}
\end{figure}

\begin{figure}[h]
\centering
\includegraphics[width=10cm, height=7cm]{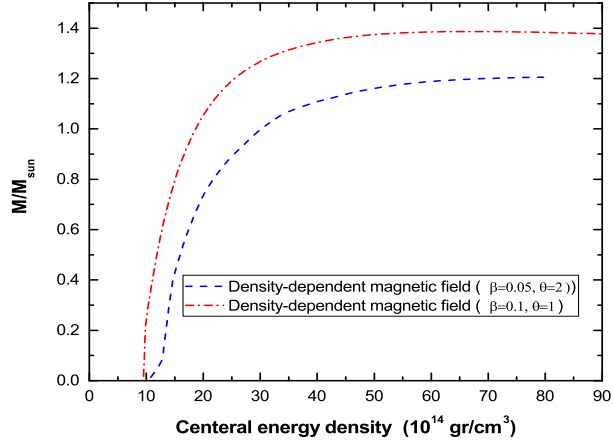}
\caption{The gravitational mass versus the central energy density of SQS in presence of two different cases for the density-dependent of the magnetic fields (Set A and Set B) using $B_{bag}^{fixed}= 90\ MeV/fm^{3}$. } \label{08}
\end{figure}

In Figs. \ref{08} and \ref{09}, the changes of the gravitational mass versus the central energy density are plotted for two cases of bag constant ($B_{bag}^{fixed}$ and $B_{bag}^{dep.}$).
In these figures, we have compared the effect of two cases of the fast (Set A) and the slow (Set B) increasing of the magnetic field from the surface to the center of SQS.
\begin{figure}[h]
\centering
\includegraphics[width=10cm, height=7cm]{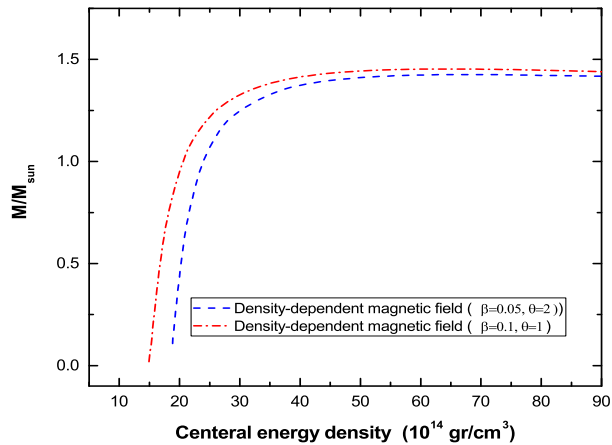}
\caption{The gravitational mass versus the central energy density of SQS in presence of two different cases for the density-dependent of the magnetic fields (Set A and Set B) using $B_{bag}^{dep.}$. } \label{09}
\end{figure}
The figures show that considering the slow changes of the magnetic field (using Set B)  leads to larger values of the maximum gravitational masses.
\begin{figure}[h]
\centering
\includegraphics[width=10cm, height=7cm]{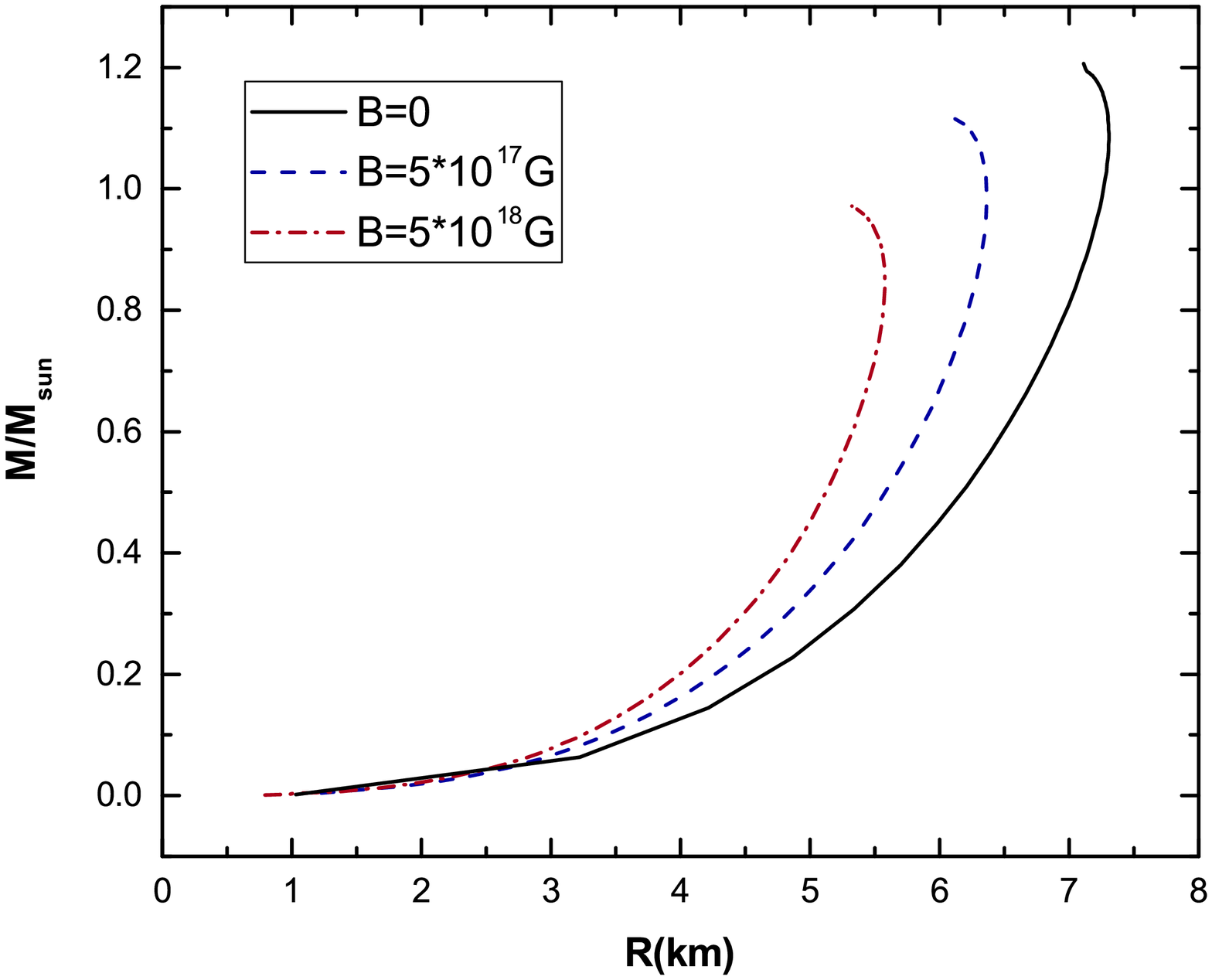}
\caption{The gravitational mass versus the radius of SQS in presence of the different magnetic fields using $B_{bag}^{fixed}= 90\ MeV/fm^{3}$. } \label{6}
\end{figure}
\begin{figure}[h]
\centering
\includegraphics[width=10cm, height=7cm]{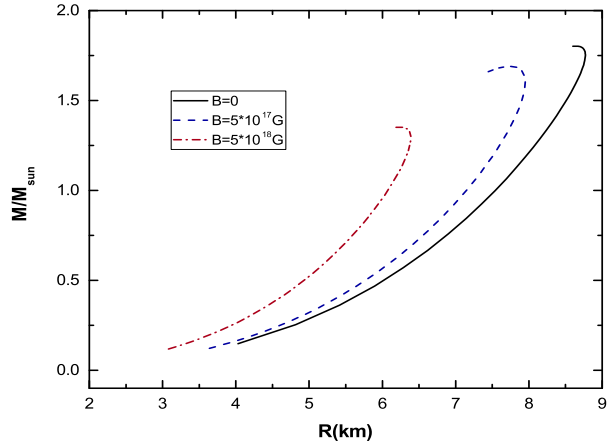}
\caption{The gravitational mass versus the radius of SQS in presence of the different magnetic fields using $B_{bag}^{dep.}$. } \label{7}
\end{figure}
\begin{figure}[h]
\centering
\includegraphics[width=10cm, height=7cm]{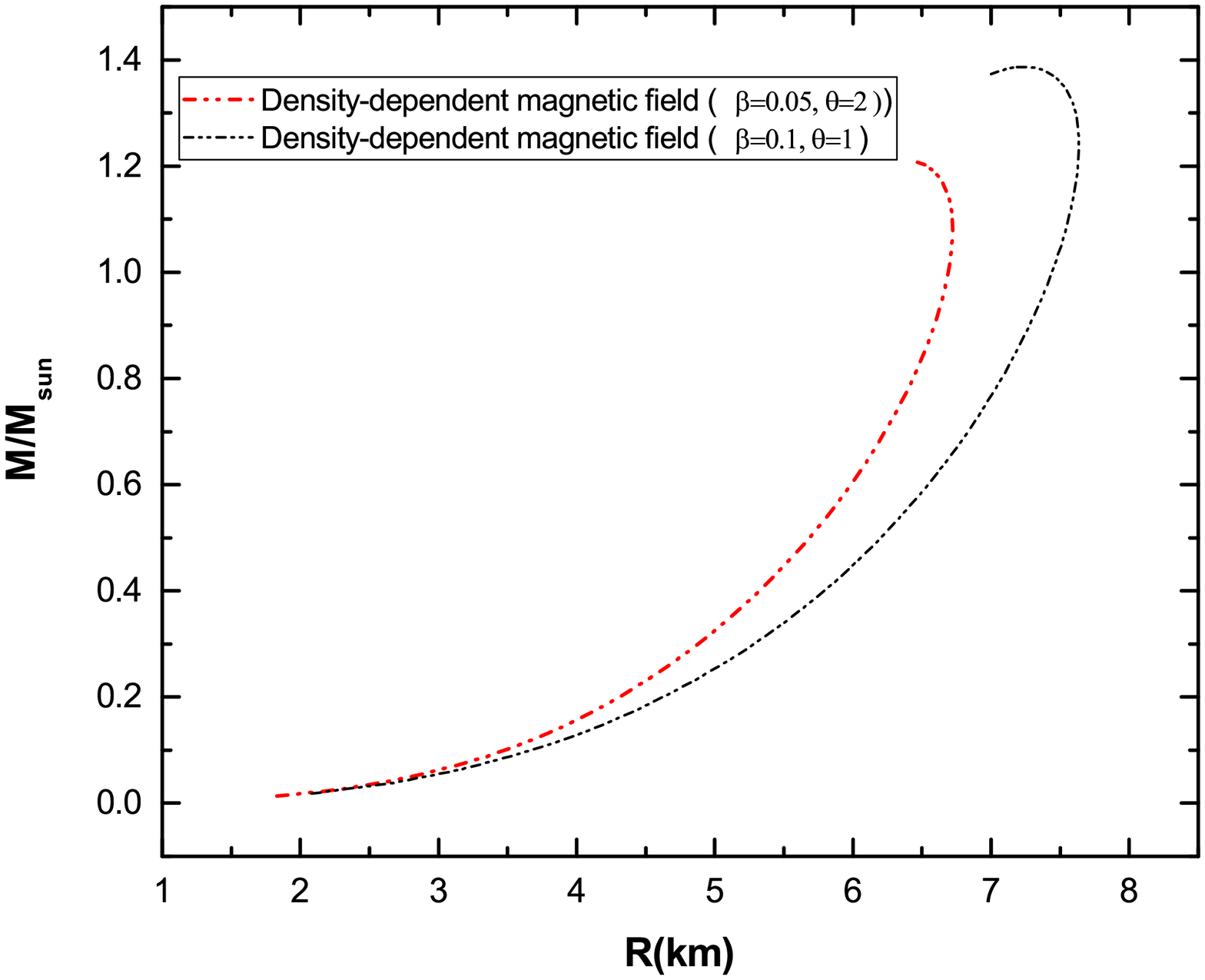}
\caption{The gravitational mass versus the radius of SQS in presence of two different cases for the density-dependent of the magnetic fields (Set A and Set B) $B_{bag}^{fixed}= 90\ MeV/fm^{3}$. } \label{8}
\end{figure}
\begin{figure}[ht]
\centering
\includegraphics[width=10cm, height=7cm]{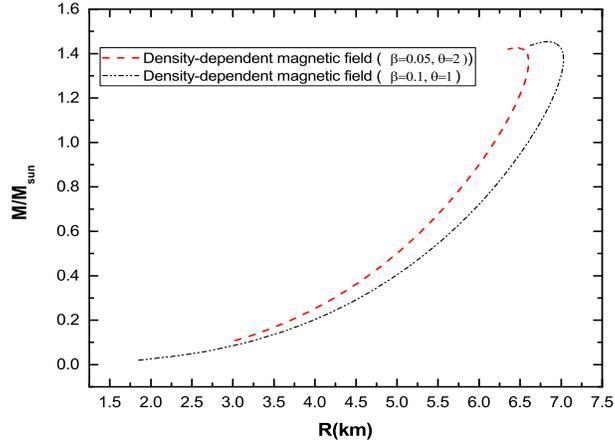}
\caption{The gravitational mass versus the radius of SQS in presence of two different cases for the density-dependent of the magnetic fields (Set A and Set B) $B_{bag}^{dep.}$. } \label{9}
\end{figure}

We have plotted the gravitational mass versus the radius of SQS for the uniform magnetic fields in Figs. \ref{6} and \ref{7} and for the density-dependent magnetic fields in Figs. \ref{8} and \ref{9},
in the both cases of $B_{bag}^{fixed}$ and $B_{bag}^{dep.}$, respectively. The results of calculation regarding SQS shows a self-bound object, in contrast to the neutron star which is bounded by gravity.
The mass-radius relation of SQS follows $M \propto R^{\alpha}$ ($\alpha \sim 3$) \cite{rk0}.
We have found that the radius increases until the gravitational mass reaches the maximum value, then the radius decreases by increasing the gravitational mass.
In other words, increasing the gravitational mass more than a certain value causes the SQS to collapses and the singularity is formed.
Also, it is clear from  Figs. \ref{6} and \ref{7}, the maximum gravitational mass and the radius decrease by increasing the magnetic field in all considered cases.
Furthermore,  Figs. \ref{8} and \ref{9} show that considering Set B for the density-dependent magnetic field leads to larger values of the maximum gravitational mass and the radius.
In addition, the $M-R$ plot corresponding to Set A (the fast drop of the magnetic field from the center to the surface of SQS) increases with the more steep slope in comparison to the same plot of Set B.
We have found with the curve fitting of Fig. \ref{8} that $\alpha$ is equal to $3.65$ for Set A and $3.41$ for Set B. Also, using Fig. \ref{9}, $\alpha$ is found $3.21$ and $3.20$ for Set A and Set B, respectively.
These values indicate that the slow changes of the magnetic field results in the smaller value of $\alpha$.

The numerical results of the structure parameters of the SQS are shown in tables \ref{T3} and \ref{T4} for two cases of bag constant, $B_{bag}^{fixed}$ and $B_{bag}^{dep.}$, respectively.
We have shown the values of compactification factor ($2M/R$) and the corresponding surface redshift ($Z_{s}=[1-2\frac{GM}{c^{2}R}]^{-\frac{1}{2}}-1$)
for the different uniform magnetic fields and $Set A$ and $Set B$ for the density-dependent magnetic field in the two latest columns.
It can be seen that the Buchdahl condition ($2M/R \leq 8/9$) is established\cite{rk53} for all considered cases. Furthermore, the value of $Z_{s}$ is in a range of $0.42 - 0.71$ that are within acceptable limits for compact objects \cite{rk54, rk55, rk56, rk57}.
Also, the maximum value of $Z_{s}$ in our calculations (for the case of density-dependent bag constant using Set B) is about $16.47\%$ lower than the upper bound of that for the subluminal EOS ($Z_{s}^{CL}=0.85$) \cite{rk057}.
As expected, the values of the surface redshift of SQS is larger than that for the results of neutron stars \cite{rk58}. The surface redshift of a neutron star is reported in a range of $0.49$ - $0.56$ in \cite{rk59}.
Also, $Z_{s}=0.51$ is calculated for a magnetic neutron star in \cite{rk60}.
The results in tables \ref{T3} and \ref{T4} show that the compactification factor increases by increasing the magnetic field.
{By investigation of the corresponding figures related to  EOS for the uniform magnetic fields (Figs. \ref{01} and \ref{02}), we have found that the stiffer EOS leads to the star with lower gravitational mass.
This behavior is different for the neutron stars where the stiffer EOS leads to the higher gravitational mass \cite{rk60a}. Interaction behavior of quarks in the strange quark matter differs from nucleon-nucleon interaction behavior in nuclear matter within the neutron star.
In addition, it is found that the stiffer EOS leads to the more compact SQS and as a result the larger value for the surface redshift.}
Furthermore, the structure parameters results for Set A and Set B in tables \ref{T3} and \ref{T4} show that the slower increase of the magnetic field from the surface to the center of SQS leads to the larger value of the compactification factor and the surface redshift.
It should be noted that in the same range of the number density, the central magnetic field reach $\simeq4.9 \times 10^{18} G$ using Set A and $\simeq3.01 \times 10^{18}\ G$
using Set B, so using Set B the EOS is softer as it can be seen from Figs. \ref{04} and \ref{05}.
In addition, by comparing the results of tables \ref{T3} and \ref{T4} the values of compactification factor in the case of $B_{bag}^{dep.}$ is larger than that for $B_{bag}^{fixed}$ with the same condition of the magnetic field.
In other words, considering the density-dependent bag constant leads to the more compact SQS.
It can be seen from tables \ref{T3} and \ref{T4} that the maximum value of compactification factor and the corresponding surface redshift are $0.89$ and $0.71$ regarding Set B of $B_{bag}^{dep.}$.
%
%
\begin{table}[ht]
\begin{center}
  \caption[]{The structure parameters of SQS in presence of different uniform magnetic fields and two different cases for the density-dependent of the magnetic field (Set A and Set B) by considering $B_{bag}^{fixed}= 90\ MeV/fm^{3}$ .}\label{T3}
  \begin{tabular}{clclclclclclclcl}
  \hline\noalign{\smallskip}
$Magnetic field$ &&  $\frac{M}{M_{sun}}$ && $R (km)$ && $\frac{2M}{R}$ &&& $Z_{s}$ &&\\
 \hline\noalign{\smallskip}
 $0$ && 1.22 && 7.10 && $0.68$  &&& $0.42$ && \\
 $5 \times 10^{17}\ G$ && 1.11 && 6.06&& $0.72$ &&& $0.47$ &&\\
 $5 \times 10^{18}\ G$ &&  0.97 && 5.32&& $0.73$ &&& $0.47$ &&  \\
 Set A &&  1.21 && 6.40 && $0.75$ &&& $0.50$ && \\
 Set B && 1.37 && 6.99 && $0.78$ &&& $0.54$ &&  \\
  \noalign{\smallskip}\hline
  \end{tabular}
\end{center}
\end{table}
\begin{table}[ht]
\begin{center}
  \caption[]{The structure parameters of SQS in presence of different uniform magnetic fields and two different cases for the density-dependent of the magnetic field (Set A and Set B) by considering $B_{bag}^{dep.}$.}\label{T4}
  \begin{tabular}{clclclclclclclcl}
  \hline\noalign{\smallskip}
$Magnetic field$ && $\frac{M}{M_{sun}}$ && $R\ (km)$ && $\frac{2M}{R}$ &&& $Z_{s}$ && \\
 \hline\noalign{\smallskip}
 $0$ && 1.80 && 8.65 && $0.82$ &&& $0.61$ && \\
 $5 \times 10^{17}\ G$ && 1.65 && 7.43&& $0.88$ &&& $0.70$ && \\
 $5 \times 10^{18}\ G$ && 1.42 && 6.37&& $0.88$ &&& $0.71$ && \\
 Set A && 1.41 && 6.33 && $0.88$ &&& $0.70$ &&  \\
 Set B && 1.44 && 6.45 && $0.89$ &&& $0.71$ &&  \\
  \noalign{\smallskip}\hline
  \end{tabular}
\end{center}
\end{table}
The structure parameters of some SQS candidates are shown in table \ref{T5}.
As it can be seen the value of the maximum gravitational mass of candidates is in the range of $1.10$ - $1.74$ $M_{sun}$,
the compactification factor is between $0.57$ - $0.83$ and the surface redshift changes from $0.31$ to $0.62$.
By comparing the values of surface redshift and the compactification factors in three tables we have found that the structure parameters of our model in all cases are in the range of those for the SQS candidates.
As our calculation results in table \ref{T4} are shown, the values of $2M/R$ and $Z_{s}$ in the case of $B_{bag}^{dep.}$ (especially using Set A and Set B)
are nearer to that values for $SAX J 1808.4 - 3659$. Actually, the comparison shows that the SQS model in the current paper is compatible with $SAX J 1808.4 - 3659$.
In addition, the values of $2M/R$ and $Z_{s}$  for $4U 1608 - 52 $ and $4U 1820 - 30 $ is comparable with the results of table \ref{T3}.
Also, the compactification factor of $RX J185635-3754$ and $Her X - 1$ are $0.60$ and $0.57$, respectively, and are comparable with the results of
 table \ref{T3} that are in the range of $0.68$-$0.78$.
\begin{table}[h]
\begin{center}
  \caption[]{Structure parameters of strange quark stars candidates.}\label{T5}
  \begin{tabular}{clclclclclclclcl}
  \hline\noalign{\smallskip}
Observed stars && $\frac{M}{M_{sun}}$ && $R (km)$ && $\frac{2M}{R}$ &&& $Z_{s}$ &&\\
 \hline\noalign{\smallskip}

$RX J185635-3754$ \cite{rk1} && 1.20 && 8.00 && $0.60$ &&& $0.34$ &&  \\
$Her X - 1$ \cite{rk62} && 1.10 && 7.70 && $0.57$ &&& $0.31$ &&\\
$4U 1608 - 52 $ \cite{rk63} && 1.74 && 9.3 && $0.74$ &&& $0.49$ &&  \\
$4U 1820 - 30 $ \cite{rk64} && 1.58 && 9.1 && $0.69$ &&& $0.43$ &&  \\
$SAX J 1808.4 - 3659 (SS1)$ \cite{rk65} && 1.44&& 7.07&& $0.80$ &&& $0.57$ && \\
$SAX J 1808.4 - 3659 (SS2)$ \cite{rk65} && 1.32 && 6.35&& $0.83$ &&& $0.62$ &&  \\

  \noalign{\smallskip}\hline
  \end{tabular}
\end{center}
\end{table}
\section{Conclusion}

In the current paper, we have investigated the thermodynamic properties of strange quark matter in the core of strange quark stars.
We have calculated the equation of state of strange quark matter using the MIT bag model in the presence of the strong magnetic field.
We have considered uniform magnetic fields, the density-dependent magnetic field and two cases of the bag constant (fixed bag constant and the density-dependent bag constant).
It is shown that the equation of state of the strange quark matter becomes stiffer by increasing the magnetic field in all considered cases.
Also, we have investigated the energy conditions based on the limitation in the energy-momentum tensor $T^{\mu \nu}$. We have found that the energy conditions are satisfied regarding the equation of state for all considered cases in this paper.
Next, we have calculated the structure parameters of the strange quark star using the Tolman-Oppenheimer-Volkov equations. We have plotted the gravitational mass versus the central energy density of the strange quark star.
It is shown that the maximum gravitational mass decreases by increasing the magnetic field.
Also, in the case of the density-dependent magnetic field, the maximum gravitational mass has a larger value when the magnetic field increases faster from the surface to the center of the star.
We have plotted the gravitational mass versus the radius of the star in the all considered cases and it is shown that the maximum gravitational mass and the radius decrease by increasing the magnetic field.
In addition, we have calculated the mass-radius relation of the strange quark star in the density-dependent magnetic field case.
We have found that $\alpha$ in the mass-radius relation of the strange quark stars ($M \propto R^{\alpha}$) is in the range of the expected values ($\alpha \sim 3$).
Also, the $\alpha$ value is larger in the case of the fast drop of the magnetic field from the center to the surface of the star.
Finally, the compactification factor and the surface redshift of the star are calculated in all considered cases. As the results have shown, the strange quark star becomes more compact by increasing the magnetic field.
Also, it becomes a more compact star when magnetic field has a slower rate of changes interior of the star. Furthermore, by considering the density-dependent bag constant in comparison with the fixed bag constant the star is denser.
It is found that the surface redshift values are in the ranges of that of the strange stars candidates. furthermore, as it is expected the larger compactification factor results in the larger redshift value.
As the results show the surface redshift and the compactification factor of the strange quark star under the conditions in this paper is comparable with the reported values of the strange quark stars candidates, considered in this paper.

\section*{Acknowledgements}
{G. H. Bordbar wishes to thank Shiraz University
Research Council.}

\end{document}